\documentstyle[12pt]{article}

\newcommand{\k}{{\bf k}}
\newcommand{\q}{{\bf q}}
\newcommand{\Q}{{\bf Q}}

\newcommand\up{\uparrow}
\newcommand\down{\downarrow}
\newcommand{\lspin}{{\bf S}}
\newcommand{\coskx}{\cos  k_{x}a}
\newcommand{\cosky}{\cos  k_{y}a}
\newcommand{\sinkx}{\sin  k_{x}a}
\newcommand{\sinky}{\sin  k_{y}a}
\newcommand{\cosqx}{\cos  q_{x}a}
\newcommand{\cosqy}{\cos  q_{y}a}

\newcommand{\spin}{\mbox{\boldmath$\sigma$}}
\newcommand{\ds}{\displaystyle}

\newcommand{\jpsj}{J.Phys.Soc.Jpn.}

\newcommand{\prl}{ Phys. Rev. Lett.}
\newcommand{\prb}{ Phys. Rev.{\bf  B}.}
\newcommand{\pr}{ Phys. Rev.}

\begin{document}

\title{Spin Fluctuations in an Itinerant Heisenberg System: Naive RPA
Treatment}

\author{ JunIchiro KISHINE\thanks{kishine@tansei.cc.u-tokyo.ac.jp}\\
\it Institute of Physics, College of Arts and Sciences, \\
\it University of Tokyo, \\
\it Komaba 3-8-1, Meguro-ku, Tokyo, Japan}
\date{July 10, 1995}
\maketitle

\begin{abstract}

The dynamical spin fluctuations in a two-dimensional square lattice
in its paramagnetic phase are examined within the framework of Random Phase
Approximation(RPA).
Itinerant carriers with spin interact with each other
via an antiferromagnetic Heisenberg interaction.

Then there appear three fundamental scattering processes;
a) scattering between parallel spins,
b) scattering between antiparallel spins and
c) scattering with spin-flip.
To examine how these scattering processes affect the dynamical spin
fluctuations,
we pick up carefully all possible combination of RPA diagrams
in a consistent manner and take the spin rotational symmetry into account.
Then it becomes clear that we have to take up a sequence of the irreducible
single loop
which in itself is modified due to the particle-hole ladder type vertex
correction.
We set up the Bethe-Salpeter equation for the vertex correction and show that
this can be
solved in a closed form due to separable nature of the antiferromagnetic
interaction.
We evaluated numerically  the effect of the vertex correction and
found that the correction is negligibly small.

Therefore we propose that in  an itinerant Heisenberg system, including the
$t$-$J$ model as its derivertive,
the simplified RPA, where
the irreducible single loop is unrenormalized, works very well.
This conclusion strongly support the simplified treatment which is widely
used in High-$T_{c}$ problem. Moreover the present formalism enables us to
proceed further
microscopic calculations on the magnetic properties in the current High-$T_{c}$
problem.

\end{abstract}
\baselineskip 15pt
\section{Introduction}

 The purpose here is to explore the dynamical spin fluctuations in a
paramagnetic
state of a two-dimensional square lattice electron system in terms of the
itinerant
Heisenberg model
within the framework of the
naive random phase approximation (RPA).
In this model the free motion of the itinerant carrier, electron or hole with
spin,
is modified by scattering processes via an antiferromagnetic Heisenberg
interaction\footnote{Although the present work was motivated by the
usual treatment of the $t$-$J$ model.
We thought it better to aboid calling the present model "$t$-$J$" model,
since the present model neglect the non-double occupancy condition due to
the strong crrelation. Our standpoint is similar to Ref\cite{Maki} in which
the spinon loop is regarded as ordinary hole and the effect of the non-double
occupancy condition is not taken into account.}.
Then there appear  three fundamental scattering processes;
a) scattering between carriers with  parallel spins,
b) scattering between carriers with  antiparallel spins and
c) scattering between carriers with spin-flip.
These processes affect each other and appear in the renormalized spin
fluctuating
processes in all possible ways. So far, however,  little attention has been
given to
the naive treatment of this problem.

The similar problem has been studied, in connection with current
High-$T_{c}$ problems\cite{ZhangRice},  in terms of the $t$-$J$ model which can
be regarded as the derivative
of the present model.
In the $t$-$J$ model where the effect of
non-double occupancy condition due to the strong correlation is incorporated
and consequently the magnetic processes are governed by the spinon's degree of
freedom.
Then there exist  scattering processes between the free spinons via an
antiferromagnetic Heisenberg interaction. In this sence the same problem occurs
concerning
the spin fluctuating processes as in the case of the present model.

Indeed, within the framework of a
mean field picture of the $t$-$J$ model,
the spin fluctuations  have been intensively studied by Tanamoto, Kohno, and
Fukuyama (T.K.F)\cite{Fukuyama}
where the dynamical spin-spin correlation function,
$
\chi_{\rm T.K.F}(q)=\chi_{\rm spinon}[1+J_{\q}\chi_{\rm spinon}(q)]^{-1},
$
plays the central role.
Here $\chi_{\rm spinon}(q)$ is the irreducible single loop
of a free spinon and $J_{\q}=2J(\cosqx+\cosqy)$ is an antiferromagnetic
Heisenberg interaction in a momentum space.

However the form of the correlation function $\chi_{\rm T.K.F}(q)$ include only
one single scattering
channel and, from diagrammatic point of view, consists of a sequence of the
irreducible single loop
which is {\it unrenormalized}. Moreover this form of the susceptibility doesn't
contain the spin indexies
as $\chi^{\alpha\beta}$. The simple form similar to $\chi_{\rm T.K.F}(q)$ has
been widely used in the
context of the High-$T_{c}$ problem.

The present work is motivated by a {\it naive} question; "when we consider all
the possible
scattering processes within RPA, what kind of the scattering processes
contribute to the dynamical spin fluctuations
and consequently the simplified structure of $\chi_{\rm T.K.F}(q)$ should be
modified
or not?".

In the present paper we put aside the effect of the strong correlation and
concentrate only on the structure of possible scattering processes which
produce the
dynamical spin fluctuations within the framework of RPA.

Generally the {\it naive} RPA corresponds to picking up  the so-called
{\it ring diagram} which is a sequence of the {\it irreducible single loop}.
Then the irreducible single loop  naturally
includes the particle-hole exchange scattering processes as the lowest
order vertex correction\cite{AndersonRPA}. This treatment fully reproduces an
equation of motion method.

However, in specific problems, how to choose the proper diagrams is
model-dependent.
For example, in the paramagnon theory based on the Hubbard
model\cite{Paramagnon},
the transverse spin fluctuations consist of one renormalized single
 loop incorporating
the particle-hole exchange scattering  and the longitudinal spin
fluctuations consist of a sequence of the unrenormalized single loop.
This situation comes from the short range nature  of the Hubbard interaction.

As an another example, in the  Coulomb gas problem, the paramagnetic
susceptibility
contains a single loop
modified due to the exchange scattering\cite{CoulombGas}.

It will be shown that in the present case both of the transverse and
longitudianl
spin fluctuations consist of a sequence of  the irreducible single loop
modified due to the particle-hole exchange scattering processes.

Since we consider a paramagnetic phase,
it is also important to  pay our attention to the spin rotational symmetry of
the theory.

  In \S 2 we present the model. In \S 3 we review the general
  formalism for the spin fluctuation and construct the naive RPA. Then
  we show that Bethe-Salpeter equation for the vertex correction
  can be solved in a closed form due to the separable form of the
  exchange interaction.
We obtain the expression for the spin fluctuations in a closed form, which is
different from the form where the irreducible single loop is unrenormalized.
  In \S 4 we discuss the spin rotational symmetry in the present treatment.
  Finally in \S 5 we present the numerical results and discuss
  quantitative difference between the present treatment and the simlified one.

   As a result we propose that the simplified treatment for the
spin fluctuations of a square lattice which has been widely used
works fairly well in the present framework.

\section{Model}

We start from the itinerant Heisenberg model,
\begin{equation}
{\cal{H}}={\cal{H}}_{0}+{{\cal{H}}_{int}},
\label{eqn:Hamiltonian}\end{equation}

Here
\begin{equation}
{\cal{H}}_{0}
=\displaystyle{\sum_{\k\sigma}}
\xi_{\k}c_{\k\sigma}^{\dagger}
c_{\k\sigma},\end{equation}
is the kinetic Hamiltonian where
$c{_{\k\sigma}}^{\dagger}$ ($c_{\k\sigma}$)
is a creation (annihilation) operator of an itinerant carrier with the momentum
$\k$ and
the spin projection $\sigma$.

Further
\begin{equation}
{\cal H}_{int}
=2 \displaystyle{\sum_{\q}}J_{\q}
\lspin_{\q}\cdot\lspin_{-\q},
\end{equation}
represents the antiferromagnetic Heisenberg interaction between
the nearest neighbor spins where
\begin{eqnarray}
J_{\q}=J (\cosqx+\cosqy),
\end{eqnarray}
denotes the momentum dependent spin-spin
interaction.
The spin fluctuation operator with momentum $\q$ is defined by
\begin{eqnarray}
\lspin_{\q}
={1\over 2}\sum_{\k,\alpha\beta}
c^{\dagger}_{\k+\q,\alpha}{\spin}_{\alpha\beta}
c_{\k\beta}.
\end{eqnarray}
where $\spin$ denotes the usual Pauli matrixes and we set $\hbar=1$.

Bearing the high-$T_{c}$ problem in mind,
we include
not only the nearest neighbor hopping, $t$, but also   the next nearest
neighbor
hopping, $t'$, and then  we have
\begin{eqnarray}
\xi_{\k}=\varepsilon_{\k}-\mu=-2t(\coskx+\cosky-\alpha\coskx\cosky)-\mu ,
\label{eqn:band}
\end{eqnarray}
where $a$ is the lattice constant and $\alpha=-2t'/t$. Here $0\leq\alpha\leq1$
is a
parameter which characterlizes the geometry of the Fermi contour.
Furthermore $\mu$ denotes the chemical potential.

The dispersion (\ref{eqn:band}) produces
the density of states (DOS);
\begin{eqnarray}
{\cal D}_{\alpha}(\varepsilon)
={\sqrt{2}\over \pi^2}{1\over \sqrt{2+\alpha\varepsilon} }
{\bf K}[\sqrt{4-(\alpha-\varepsilon/2t)^{2}\over
2(2+\alpha\varepsilon)}]\label{eqn:DOS}.
\end{eqnarray}
where ${\bf K}(x)$ denotes the elliptic integral of the first kind. The
derivation of
(\ref{eqn:DOS}) will be presented in Appendix A. The carrier density in the
electron picture is determined by
\begin{eqnarray}
n\equiv{N_{c}\over N}=\ds \int_{-2 t (2-\alpha)}^{2 t (2+\alpha) }d\varepsilon
{\cal D}_{\alpha}(\varepsilon) f(\varepsilon),
\end{eqnarray}
where $f(\varepsilon)={(e^{\varepsilon-\mu\over T}+1)^{-1}}$. $N_{c}$ and $N$
represent
the number of carrier and lattice sites respectively. Then $n=1$ corresponds to
the half-filling.

This type of DOS shows the logarithmic van-Hove singularity at
$\varepsilon=-2\alpha$ for $0\leq\alpha<1$ and the square root singularity for
$\alpha=1$, which originates from the saddle points located at the four points
$(\pm\pi,0),(0,\pm\pi)$ in the first
Brillouin zone.
The case of $\alpha=0$ which corresponds to LSCO compounds leads to the perfect
nesting.
As the parameter $\alpha$ becomes nearer to $\alpha=1$, the
curvature of the convex along the $\Gamma$-$Y$ line near the saddle point
becomes flatter.
Then $\alpha=1$ corresponds to the perfectly flat curvature of the convex.
This type of the saddle point is called "the extended saddle point"
which produces the power singurality in DOS\cite{AbriCamp}.
In the case of YBCO,  $\alpha=0.9$ corresponds to the real data\cite{HopInte},
which is very near the case of the extended saddle point.
In Fig.~1. we show the energy contour, corresponding DOS, and the structure
of the saddle point near the Y point $(0,\pi)$ for typical values of $\alpha$.

\section{Transverse Spin Fluctuation}
\subsection{General Formalism}

 We consider the magnetic response in a square lattice electron system within
the framework of
the present model.
 Whole information on spin dynamics is contained in
the dynamical wave number dependent magnetic susceptibility,
\begin{eqnarray}
\chi{^{\alpha\beta}}(q)
&=&
\int_{0}^{\beta}d\tau e^{i\omega_{n}\tau}
<T_{\tau}[S^{\alpha}(\q,\tau)S^{\beta}(0)]>
={1\over 4} \sum_{\mu\nu\lambda\rho}
\sigma^{\alpha}_{\mu\nu}
\Gamma^{\mu\nu;\lambda\rho}(q)
\sigma^{\beta}_{\lambda\rho}.
\end{eqnarray}
where
\begin{equation}
\Gamma^{\mu\nu;\lambda\rho}(q)=
\int_{0}^{\beta}d\tau e^{i\omega_{n}\tau}
{1\over N}\sum_{\k,\k'}
<T_{\tau}[c^{\dagger}_{\k',\mu}(\tau)c_{\k'+\q,\nu}(\tau)
c^{\dagger}_{\k+\q,\rho}(0)c_{\k,\lambda}(0)]>,
\end{equation}
denotes the spin dependent polarization function
where $\alpha,\beta=+,-,z$ and
$\mu,\nu,\lambda,\rho=\uparrow,\downarrow$ and $i\omega_{n}=2\pi i T n$ is a
bosonic Matsubara frequency.
Throughout the present paper $q$ denotes a four vector $q=(\q,i\omega_{n})$.

Furthermore
\begin{eqnarray}c^{\dagger}_{\k,\alpha}(\tau)
= \exp(\tau{\cal H}) c^{\dagger}_{\k,\alpha}(0) \exp(-\tau{\cal H}),
\end{eqnarray}
represents imaginary time dependent creation operator
 where we set $\hbar=1$ and $k_{B}=1$.
$<\cdots>\equiv {\rm Tr}( e^{-{\cal H}\over T}\cdots)/{\rm Tr}{e^{-\cal H}\over
T}$
denotes the thermal average under the full Hamiltonian.

 By noting
\begin{eqnarray}
\sigma^{+}=\left( \begin{array}{cc}0&1\\0&0\end{array} \right),
\sigma^{-}=\left( \begin{array}{cc}0&0\\1&0\end{array} \right),
\sigma^{z}=\left( \begin{array}{cc}1&0\\0&-1\end{array} \right),
\end{eqnarray}
the longitudinal and transverse spin susceptibility  can be obtained as
\begin{eqnarray}
\chi^{+-}(q)&=&\Gamma^{\up\down;\down\up}(q),\\
\chi^{-+}(q)&=&\Gamma^{\down\up;\up\down}(q),\\
\chi^{zz}(q)={1\over
4}[\Gamma^{\up\up;\up\up}(q)+\Gamma^{\down\down;\down\down}(q)&-&\Gamma^{\up\up;\down\down}(q)
-\Gamma^{\down\down;\up\up}(q)]={1\over
2}[\chi^{\up\up}(q)-\chi^{\up\down}(q)].
\label{eqn:pol for zz}\end{eqnarray}
where
$\chi^{\sigma,\sigma'}(q)\equiv\Gamma^{\sigma,\sigma;\sigma',\sigma'}(q)$.

Throughout this paper we consider the paramagnetic pahse and therefore the
dynamic spin susceptibilities
must satisfy the rotational symmetry relation in spin space;
\begin{equation}
2\chi^{zz}=\chi^{+-}.\label{eqn:RotationalSymmetry}
\end{equation}

 The  dynamical susceptibility
of a free carrier system can be given by
\begin{eqnarray}
\chi_{\rm free}(q)
=-{T\over N}\sum_{k}{\cal G}_{0}(k+q) {\cal G}_{0}(k).
\label{eqn:bare poralization}
\end{eqnarray}
Here
\begin{equation}
{\cal G}_{0}(k)={1\over i\varepsilon_{n}-\xi_{\k}},
\end{equation}
is a thermal green's function of a free carrier where
$\varepsilon_{n}=(2n+1)\pi T$ is a fermionic Matsubara frequency.
Furthermore from now on the sumation over the four momentum means
\begin{equation}{1\over
N}\sum_{k}=\sum_{n=-\infty}^{\infty}\int_{-\pi}^{\pi}{dk_{x}a\over
2\pi}\int_{-\pi}^{\pi}{dk_{y}a\over 2\pi}.\end{equation}
By taking the Matsubara summation we obtain
\begin{equation}
\chi_{\rm free}(q)={1\over 2N}\sum_{\k}
{
\Lambda_{\k,\q}(T)
(\xi_{\k+\q}-\xi_{\k})
\over
\omega_{n}^{2}+(\xi_{\k+\q}-\xi_{\k})^{2}
\
},
\end{equation}
where
$\varepsilon_{n}$ is a fermionic Matsubara frequency and
\begin{equation}
\Lambda_{\k,\q}(T)=\tanh(\displaystyle{\xi{_{\k+\q}}\over 2T})-
\tanh(\displaystyle{\xi_{\k} \over 2T}),
\end{equation}
denotes the thermal extinction factor.

\subsection{RPA}

 Now we consider the dynamical susceptibiliy which is modified
due to $J_{\q}$. We calculate the dynamical susceptibility within the framework
of RPA.

The fundamental processes induced by $J_{\q}$ are produced
through
\begin{equation}
\lspin_{\q}\cdot \lspin_{-\q}={1\over
2}(S^{+}_{\q}S^{-}_{-\q}+S^{-}_{\q}S^{+}_{-\q})+S^{z}_{\q}S^{z}_{-\q}.
\label{eqn:InterlayerHeisenberg}\end{equation}
By rewriting this expression in a second quantized form, we get
three fundamental vertexes for the scattering processes between propagating
carriers with spin.

The first term of (\ref{eqn:InterlayerHeisenberg}) produces the scattering
process
with spin flip (type-a).
The second term of (\ref{eqn:InterlayerHeisenberg}) produces two fundamental
processes;
scattering processes between parallel spins (type-b) and anti-parallel spins
(type-c).
These processes are shown in Fig.~2.

 Here we first consider the transverse spin fluctuation.  The
 problem of the rotational symmetry in spin space will be taken up in the next
section.
The  string type summation as shown in Fig.~3.  produces the form,
\begin{equation}
\chi^{+-}(q)=
{\tilde\chi_{0}(q)\over 1+J_{\q}{\tilde\chi}_{0}(q)},\label{eqn:string}
\end{equation}
where ${\tilde\chi_{0}}(q)$ is the irreducible single fermion loop which cannot
be cut
into two pieces by removing an interaction line and which can be written in the
form
\begin{equation}
{\tilde\chi_{0}}(q)=-{T\over N}\sum_{k}
{\cal G}(k){\Gamma}(k,k+q){\cal G}(k+q),\label{eqn:modintra}
\end{equation}
where ${\Gamma}(k,k+q)$ is the triangle  vertex inserted in a single loop.

Furthermore
${\cal G}(k)$
is a dressed green's function determined through the Dyson equation
\begin{equation}
{\cal G}(k)^{-1}={\cal G}_{0}(k)^{-1}-\sum(k),\label{DysonEq}
\end{equation}
where
$\sum(k)$ denotes the self energy part.

We can see that in  Fig.~3.  only the type-a processes can contribute to
the ring diagram for the transverse susceptibility within RPA. Here we should
note that, even within the
framework of RPA, the type-c processes can
produce the exchange scattering between a electron and hole inside a single
loop.
Consequently we have to include the particle-hole ladder process in the vertex
$\Gamma(k+q,k)$.
These type of diagrams are characteristic to the $t$-$J$ model
while in Hubbard model the string type summation cannot appear in the
transverse fluctuation.

Therefore we have to include the string type summation and
the ladder type summation simultaneously to get the naive result within RPA.

To treat the compatibility of the self-energy and the vertex correction,
it is sufficient to consider only the Fock term
to ensure the  Ward-Takahashi identity\cite{Shastry}.
This procedure corresponds to the simplest case
of Baym-Kadanoff's {\it conserving approximation}\cite{BK}.
Then $\sum(k)$ can simply be written by
\begin{equation}
\sum(k)=-{3\over 2}{T\over N}\sum_{k'}J(\k-\k'){\cal G}(k'),
\label{eqn:SE}\end{equation}
which can be reduced to the form
\begin{equation}
\sum(k)=
-{3\over 4}J X(T)(\coskx+\cosky),
\end{equation}
where
\begin{eqnarray}
X(T)&=&\sum_{k'}{\cos k_{x}'a+\cos k_{y}'a\over i\varepsilon_{n'}
-\varepsilon_{\k'}+\mu-\Sigma(k')}\nonumber\\
&=&-{1\over 2}\displaystyle \int{d\k'\over (2\pi)^2}
(\cos k_{x}'a+\cos k_{y}'a)\tanh
[(\varepsilon_{\k'}-\mu+\Sigma(k')/2T]\label{eqn:self}.
\end{eqnarray}
Here we've performed Matsubara summmation and took into account
the ${\cal D}_{4}$ symmetry
of the system. We can obtain the self energy by solving (\ref{eqn:self})
in a self-consistent manner.
  $X(T)$ depends weakly on the temperature $T$ and turns out to have positive
value.
It follows from this situation that the Fock term, as usual,  enhances the
effective mass.
 Now the energy dispersion of a single carrier is slightly modified from
(\ref{eqn:band})
to the form
\begin{equation}
\varepsilon_{\k}=-2[t-{3\over 8} J  X(T)](\coskx+\cosky)+2\alpha\coskx\cosky.
\end{equation}

The triangle vertex ${\Gamma}(k,k+q)$  satisfies the
Bethe-Salpeter  equation for a particle-hole channel  as shown in Fig.~4.,
\begin{eqnarray}
{\Gamma}(k+q,k)=1
+{1\over 2}{T\over N}\displaystyle\sum_{k'}
J{_{\k-\k'}}{\cal G}(k'){\Gamma}(k',k'+q){\cal G}(k'+q).
\label{eqn:integral eq for the ladder vertex}
\end{eqnarray}

We can solve (\ref{eqn:integral eq for the ladder vertex}) in the closed form
as
\begin{eqnarray}
\Gamma(k,k+q)=1+
J C(q)\,\,(\coskx+\cosky) +J S(q)\,\,(\sinkx+\sinky)
\label{eqn:ladder vertex}
\end{eqnarray}
where
\begin{eqnarray}
C(q)&=& {\displaystyle{-{1\over 2}\chi_{1}(q)[1+{1\over 2}J\chi_{5}(q)]
 +J[{1\over 2}\chi_{2}(q)][{1\over 2}\chi_{4}(q)]
\over
[1+{1\over 2}J\chi_{3}(q)][1+{1\over 2}J\chi_{5}(q)]
-[{1\over 2}J\chi_{4}(q)]^{2} }}\label{eqn:simeq C},\\
S(q)&=& {\displaystyle{-{1\over 2}\chi_{2}(q)[1+{1\over 2}J\chi_{3}(q)]
 +J[{1\over 2}\chi_{1}(q)][{1\over 2}\chi_{4}(q)]
\over
[1+{1\over 2}J\chi_{3}(q)][1+{1\over 2}J\chi_{5}(q)]-
[{1\over 2}J\chi_{4}(q)]^{2}}}.
\label{eqn:simeq S}
\end{eqnarray}
We may leave the details of the derivation of this formula to Appendix B.
Here  $\chi_{1}(q)$, $\cdots$, $\chi_{5}(q)$ are defined by
\begin{eqnarray}
\begin{array}{c}
\displaystyle
{\chi}_{1}(q)=-{T\over N}\sum_{k}
(\cos k_{x}a+\cos k_{y}a){\cal G}(k'){\cal G}(k'+q),
 \\
\displaystyle
\chi_{2}(q)=-{T\over N}\sum_{k}
(\sin k_{x}a+\sin k_{y}a){\cal G}(k){\cal G}(k+q),
\\
\displaystyle
\chi_{3}(q)=-{T\over N}\sum_{k}
(\cos k_{x}a+\cos k_{y}a)^{2}{\cal G}(k){\cal G}(k+q),
\\
\displaystyle
\chi_{4}(q)=-{T\over N}\sum_{k}
(\cos k_{x}a+\cos k_{y}a)
\,\,\,\,\,\,\,\,\,\,\,\,\,\,\,\,\,
\\(\sin k_{x}a+\sin k_{y}a){\cal G}(k){\cal G}(k+q),
\\
\displaystyle
\chi_{5}(q)=-{T\over N}\sum_{k}
(\sin k_{x}a+\sin k_{y}a)^{2}{\cal G}(k){\cal G}(k+q).
\end{array}\label{eqn:weighted bubble}
\end{eqnarray}

Therefore we can obtain the result
\begin{equation}
{\tilde\chi}_{0}(q)
={\chi}_{0}(q)+JC(q){\chi}_{1}(q)+JS(q){\chi}_{2}(q).
\label{eqn:finalresult}\end{equation}
where
\begin{equation}
\chi_{0}(q)=-{T\over N}\sum_{k}
{\cal G}(k+q){\cal G}(k).
\end{equation}

We note here that for the commensurate spin-fluctuation with
$\q=\Q=(\pi/a,\pi/a)$, $\chi_{2}=\chi_{4}=0$ due to the  $D_{4}$ symmetry of
the
square lattice, and then consequently $S(q)=0$.

Directly from the above expressions (\ref{eqn:weighted bubble}),
we can expect the correction terms in the (\ref{eqn:finalresult})
are negligiblly small. Therefore we expect in our scheme the vertex correction
coming from the exchange scattering processes can be neglected.
We will confirm this situation numerically in \S 5.

\section{Rotational Symmetry in Spin Space}

When we consider the spin fluctuations in a paramagnetic phase, to ensure the
consistency of the theory, the
dynamical spin susceptibilities have to satisfy the symmetry relation
(\ref{eqn:RotationalSymmetry})
in  spin space.
Here we prove the rotational symmetry of the theory in the present scheme.
Our concern is to show the relation (\ref{eqn:RotationalSymmetry}),
\begin{equation}
2\chi^{zz}=\chi^{+-}\label{eqn:spin rotational symmetry}.
\end{equation}

First we show that the spin rotational  symmetry is satisfied in a single  loop
level.
Namely
\begin{equation}
2\tilde\chi_{0}^{zz}=\tilde\chi_{0}^{+-}\equiv
\tilde\chi_{0}\label{eqn:singlelevel},
\end{equation}
where $\tilde\chi_{0}$ was given by (\ref{eqn:modintra}).
Then we have to include various intermediate spin configurations between a
given initital and final
spin configuration.
We consider a single loop of the $n$-th order term, $\tilde\chi_{0
(n)}^{\sigma\sigma'}$, with respect to $J_{\q}$.
There can appear the type-a and the type-b vertexes in all the possible manner,
as is shown in Fig.~5.
We have to take summation with respect to all the possible intermediate spin
configurations and pick up a lot of diagrams.
Fortunately we can obtain the relation between
$\tilde\chi{_{0}}{^{\up\up}}_{(n)}$,
$\tilde\chi_{0 (n)}^{\up\down}$ and $\tilde\chi_{0 (n)}^{+-}$ by order-by-order
consideration.
The results are as follows.

When $n=2m$,
\begin{eqnarray}
\tilde\chi_{0 (2m)}^{\up\up}&=&\tilde\chi_{0
(2m)}^{+-}\sum_{k=0}^{m}2^{2k}{_{2m}}C_{2k},\label{eqn:ex}\\
\tilde\chi_{0 (2m)}^{\up\down}&=&\tilde\chi_{0
(2m)}^{+-}\sum_{k=0}^{m-1}2^{2k+1}{_{2m}}C_{2k+1}.
\end{eqnarray}

When $n=2m+1$,
\begin{eqnarray}
\tilde\chi_{0 (2m+1)}^{\up\up}&=&
-\tilde\chi_{0 (2m+1)}^{+-}\sum_{k=0}^{m}2^{2k}{_{2m+1}}C_{2k},\\
\tilde\chi_{0 (2m+1)}^{\up\down}&=&
-\tilde\chi_{0 (2m+1)}^{+-}\sum_{k=0}^{m-1}2^{2k+1}{_{2m+1}}C_{2k+1},
\end{eqnarray}
where $_{k}C_{l}={k!/ [(k-l)!l!]}$ is a binomial coefficient.
Here, for example,  we derive the formula (\ref{eqn:ex}).
In this case $n=2m$ and thererfore the single loop can include
the even number of the type-a processses. Then all the other
vertexes are the type-b vertexes.
We consider the case when there are $2k$ type-a processses and
$2m-2k$ type-b vertexes.
Since the type-a vertex gives the facor $J_{\q}$ and
the type-b vertex gives the facor $J_{\q}/2$, if we replace all the
type-a) vertexes simply by the type-b vertexes, there appear the factor
$2^{2k}$. On the other hand there are $_{2m}C_{2k}$ ways of locating the type-a
vertexes inside a loop. As a result the corresponding expression has a factor
$2^{2k}{_{2m}}C_{2k}$ and therefore we obtain (\ref{eqn:ex}).
Here $\tilde\chi_{0 (n)}^{+-}$ is a single loop corresponding to the transverse
fluctuation of the same order with respect to $J_{\q}$.
Now by noting that
$$\sum_{k=0}^{m}2^{2k}{_{2m}}C_{2k}-\sum_{k=0}^{m-1}2^{2k+1}{_{2m}}C_{2k+1}
=(1-2)^{2m}=1,$$ and
$$\sum_{k=0}^{m}2^{2k}{_{2m+1}}C_{2k}-\sum_{k=0}^{m-1}2^{2k+1}{_{2m+1}}C_{2k+1}
=(1-2)^{2m+1}=-1,$$
we can see that for arbitraly order,
\begin{equation}
\tilde\chi_{0 (n)}^{\up\up}-\tilde\chi_{0 (n)}^{\up\down}=\tilde\chi_{0
(n)}^{+-},
\end{equation}
and consequently
\begin{equation}
\tilde\chi_{0}^{\up\up}-\tilde\chi_{0}^{\up\down}=\tilde\chi_{0}^{+-}=\tilde\chi_{0}.
\label{rotinvinsingle}\end{equation}
Therefore we obtain the  result (\ref{eqn:singlelevel}) which would be
expected.

Next we consider the string type series for the longitudinal spin fluctuations.
In the string processes, in this case,  there can appear the type-b and type-c
vertexes.
Then
the $m$-th order term with respect to $J_{\q}$ can be expressed by
\begin{eqnarray}
C^{\sigma\sigma'}_{m}=(-{J_{\q} \over 2})^{m}
\sum_{\sigma_{1}, \sigma_{2}, \cdots,
\sigma_{2m}}(-1)^{\sigma_{1}-\sigma_{2}}(-1)^{\sigma_{3}-\sigma_{4}}\cdots
(-1)^{\sigma_{2m-1}-\sigma_{2k}}\\
\tilde\chi_{0}^{\sigma\sigma_{1}}\tilde\chi_{0}^{\sigma_{2}\sigma_{3}}\cdots
\tilde\chi_{0}^{\sigma_{2m}\sigma'}.
\label{eqn:n-th C}
\end{eqnarray}
This process is shown in Fig.~6.

By using (\ref{eqn:pol for zz}) we can obtain the longitudinal spin
susceptibility  as
\begin{equation}
\chi^{zz}={1\over 2}\sum^{\infty}_{m=0}(C^{\up\up}_{m}-C^{\up\down}_{m}).
\end{equation}

First we take the summation over $\sigma_{1}$ and $\sigma_{2}$ in
(\ref{eqn:n-th C})
to get
\begin{eqnarray}
C^{\sigma\sigma'}_{m}&=&
(-{J_{\q} \over 2})^{m}
\sum_{\sigma_{3}, \sigma_{4}, \cdots,
\sigma_{2m}}(-1)^{\sigma_{3}-\sigma_{4}}\cdots
(-1)^{\sigma_{2m-1}-\sigma_{2m}}\nonumber\\
& & \,\,\,\,\,\,\,\,\,\,\,\, \,\,\,\,\,\,\,\,\,\,\,\,\,\,\,\,\,\,\,\,\,\,\,\,
(\tilde\chi_{0}^{\sigma\sigma}-\tilde\chi_{0}^{\sigma,-\sigma})
(\tilde\chi_{0}^{\sigma\sigma_{3}}-\tilde\chi_{0}^{-\sigma\sigma_{3}})
\tilde\chi_{0}^{\sigma_{4}\sigma_{5}}\cdots
\tilde\chi_{0}^{\sigma_{2m}\sigma'}\nonumber\\
&=&
(-{J_{\q} \over 2})^{m}
\sum_{\sigma_{3}, \sigma_{4}, \cdots,
\sigma_{2m}}(-1)^{\sigma_{3}-\sigma_{4}}\cdots
(-1)^{\sigma_{2m-1}-\sigma_{2m}}\\
& & \,\,\,\,\,\,\,\,\,\,\,\, \,\,\,\,\,\,\,\,\,\,\,\,\,\,\,\,\,\,\,\,\,\,\,\,
\tilde\chi_{0}(\tilde\chi_{0}^{\sigma\sigma_{3}}-\tilde\chi_{0}^{-\sigma\sigma_{3}})
\tilde\chi_{0}^{\sigma_{4}\sigma_{5}}\cdots
\tilde\chi_{0}^{\sigma_{2k}\sigma'}.\nonumber
\label{eqn: sumover12}
\end{eqnarray}
Here we used the relation (\ref{rotinvinsingle}).

By repeating the summation procedures over the pair
$(\sigma_{2i-1},\sigma_{2i})$, we get the result for the $i$-th procedure as
\begin{eqnarray}
C^{\sigma\sigma'}_{m}=
(-{J_{\q} \over 2})^{m}2^{i}\tilde\chi_{0}^{i+1}
\sum_{\sigma_{2i+1}, \sigma_{2i+2}, \cdots,
\sigma_{2m}}(-1)^{\sigma_{2i+1}-\sigma_{2i+2}}\cdots
(-1)^{\sigma_{2m-1}-\sigma_{2m}}\nonumber\\
(\tilde\chi_{0}^{\sigma\sigma_{2i+1}}-\tilde\chi_{0}^{-\sigma\sigma_{2i+1}})\cdots \tilde\chi_{0}^{\sigma_{2m}\sigma'}.
\label{eqn: sumoveri}
\end{eqnarray}
Finally we obtain the relation
\begin{eqnarray}
C^{\sigma\sigma'}_{m}=
(-J_{\q})^{m}\tilde\chi_{0}^{m}
(\tilde\chi_{0}^{\sigma\sigma'}-\tilde\chi_{0}^{-\sigma,\sigma'}).
\label{eqn: result for k-th}
\end{eqnarray}

Therefore we get
\begin{equation}
\chi^{zz}={1\over 2}\sum^{\infty}_{m=0}(C^{\up\up}_{m}-C^{\up\down}_{m})
={1\over 2}\sum^{\infty}_{k=0}(-J_{\q})^{m}\tilde\chi_{0}^{m+1}
={1\over 2}{ \tilde\chi_{0}\over 1+J_{\q}\tilde\chi_{0}}.
\end{equation}
which is just one half of the result for the corresponding transverse
susceptibility
(\ref{eqn:string}).
Thus we have proven the expected relation (\ref{eqn:spin rotational symmetry}).

\section{ Numerical Results and Conclusion}

 In this section  we show numerically that the correction terms in
(\ref{eqn:finalresult})
 can be neglected  and justify the
anticipation that a  simplified treatment where the irreducible single loop is
unrenormalized
works  well.

For this purpose we concentrate on evaluation  of the terms
$J C(q) \chi_{1}(q)$ and $J S(q) \chi_{2}(q)$ in (\ref{eqn:finalresult}).

{}From now on, to proceed numerical work,  we set the parameters   $J=t/4$.

\subsection*{1.\it\underline{Self Energy}}
First we present the numerical results of the self-consistent equation for
the self energy; (\ref{eqn:self}).
We performed the integration over the first Brillouin zone divided into
$64\times64$
uniform mesh. We set the carrier density $n=0.85$.
In Fig.~7(a) we show the temperature dependence of the factor $X(T)$ in
(\ref{eqn:self})
for $\alpha=0$ and $\alpha=1$. We can see that the self energy depends very
weakly on
the temperature over the wide
range $0< T<200$[K]\footnote{Here we put $t\sim 130$ [meV]}.

Next in Fig.~7(b) we show the carrier density dependence of $X(T)$.
It follows from this that in case of  $\alpha=0$ the effective mass becomes
monotonically heavier as the system approaches the perfect nesting.
On the other hand in case of $\alpha=1$ the effective mass has the highest
value
near the carrie density $n\sim 0.7$

In any event  the Fock term gives rise to the negative correction to the
hopping integral
$t$ of the order ${3\over8}JX(T)\sim 0.005 t$ which can safely be neglected.
It is true that we should take the Fock term into account to ensure
the consistency of the theory, but it becomes negligiblly small in magnitude
and
can be neglected.

\subsection*{2.\it\underline{Irreducible Single Loop $\tilde\chi_{0}(q)$}}

 Next we are concerned with the correction terms in the single loop.
If the single loop can well be approximated  by the unrenormalized one,
$\chi_{0}(q)$,
we can replace the irreducible polarization by the polarization for
the free carrier system (\ref{eqn:bare poralization}), since as shown above the
self-energy correction can be neglected\footnote{
We note that if we descard the vertex
correction, it is no longer necessary  to include the self-energy correction to
ensure the Ward-Takahashi identity.}.
In such a case our treatment reproduces the effective RPA in
Ref.\cite{Fukuyama}.

Here we fix the temperature to $T=0.04 t\sim 60$[K] and take the corresponding
lowest matsubara
frequency $\omega_{1}\sim 32$[meV].
The integration is performed over the first Brillouin zone divided into
$128\times128$
uniform mesh.

First in Fig.~8 we present the numerical results for the modified polarizations
(\ref{eqn:weighted bubble}) in the first Brillouin zone for the fixed carrier
density $n=0.75$.
Fig.~8(a), (b) correspond to the cases of $\alpha=0$ and $\alpha=1$
respectively.

In case of $\alpha=0$, $\chi_{0}$, $\chi_{3}$ and $\chi_{5}$ exhibit the well
known
incommensurate peaks around the nesting vector $\q=(\pi,\pi)$, since these two
terms directly reflect the symmetry of the Fermi contour.
On the other hands  $\chi_{2}$, and  $\chi_{4}$ vanish at
$\q=(\pi,\pi)$, as was  suggested in \S3.
The reason why $\chi_{1}$ also vanish at
$\q=(\pi,\pi)$ is that then the integrand of the $\chi_{1}$  becomes odd
function
with respect to the energy.

In case of $\alpha=1$, the overall structures and magnitudes of the modified
polarizations
are similar to the case of $\alpha=0$.

It should be noted that in any case $\chi_{1}$ and $\chi_{4}$ which appear in
the
numerators of $C(q)$ and $S(q)$ in the expressions (\ref{eqn:simeq C}) and
(\ref{eqn:simeq S}) become very small
in comparison with the $\chi_{3}$ and $\chi_{5}$ which appear in the
denominators. It follows from this observation that we expect  the
terms $JC\chi_{1}$ and $JS\chi_{2}$ to become negligiblly small
in comparison with $\chi_{0}$.

Indeed as shown in Fig.~9, $JC\chi_{1}$ and $JS\chi_{2}$
have very small value in the whole region of the first Brillouin zone.
However we can say with certain that these terms have tendency to
reduce the magnetic fluctuation, because these terms have negative values
in the whole region.

We can see directly from the results shown in Fig.~9 that in case of
$\alpha=0$,
the nesting condition strongly surpresses both of $JC\chi_{1}$ and
$JS\chi_{2}$.
This means the exchange scattering processes are strongly surpressed due to
the nesting condition.

On the other hand, in case of $\alpha=1$, the exchange scattering is enhanced
as the system becomes nearer to the half-filling. Furthermore we note that
the absolute vale of these terms are smaller than that in case of $\alpha=0$.
We can say this difference comes from  the different geometry of the Fermi
contour.

Finally we compare $\chi_{0}$ and
$\tilde\chi_{0}=\chi_{0}+JC\chi_{1}+JS\chi_{2}$.
As shown in Fig.~10, we can see that we can't distinguish these two quantities.
This situation is just what we expect and therefore we can certainly say
that we can work well only by taking  $\chi_{0}$ into account.
This means in the present model, the spin fluctuations of the system can fairly
well be
described in terms of the simplified, or effective, RPA, which include
only the string type procesesses in Fig.~3 in which the single loop can be
replaced by the bare one.

We can expect that this situation can survive  in case of the $t$-$J$ model.
In the $t$-$J$ model the free carrier in the present work is repaced by the
free spinon and,
in the mean field level,
the bare hopping parameter $t$ and $t'$ are replaced by $\tilde t$ and $\tilde
t'$ which depend on
carrier concentration. However fundamental scattering processes are completely
the same as in
the present work.

In the present paper we have establised the consistent formalism
to treat the spin fluctuations  by an itinerant Heisenberg model within the
framework of RPA.
Here the conservation low of the spin and the rotational symmetly of the system
was carefully treated. As a result of these careful consideration, the
simplified version
of RPA turns out to work fairly well even in the real High-$T_{c}$ problems.

\section*{Acknowledgements}

The author  thanks Prof.H.Namaizawa for useful dicussion.
I would like to express his gratitude to Mr.H.Morishita for his valuable
comments.
The author expresses his gratitude to Minako Kishine for her
reading of the manuscript.

\section*{Appendix A}
\subsection*{Density of States}

Here we derive an analytic expression for the density of states per spin.
\begin{eqnarray}
{\cal{D}}_{\alpha}(\varepsilon)
&=&
a^2\displaystyle
\int_{-\pi/a}^{\pi/a}{dk_{x}\over 2 \pi}\int_{-\pi/a}^{\pi/a}{dk_{y}\over 2
\pi}
\delta(\varepsilon-\varepsilon_{\k})\nonumber\\
&=&
{1\over \pi^2 }\int_{0}^{\pi}dx\int_{0}^{\pi}dy
\delta(\varepsilon+2t(\cos x+\cos y -\alpha\cos x\cos y))\nonumber\\
&=&
{1\over \pi ^2 t}
\int_{-1}^{1}{d\xi\over \sqrt{1-\xi^2}}
\int_{-1}^{1}{d\eta\over\sqrt{1-\eta^2}}
\delta[\varepsilon/t-2(\xi+\eta-\alpha\xi\eta)]\nonumber\\
&=&{1\over 2\pi^2}
\int_{-1}^{1-\varepsilon/2t\over 1+\alpha}
{d\xi\over \sqrt{ (1-\xi^2)\{(1+\varepsilon/2)+(1-\alpha)\xi\}
                \{ (1-\varepsilon/2)-(1+\alpha)\xi\} } }\nonumber\\
&=&{\sqrt{2}\over \pi^2}{1\over \sqrt{2+\alpha\varepsilon} }
{\bf K}[\sqrt{4-(\alpha-\varepsilon/2t)^{2}\over
2(2+\alpha\varepsilon)}].\nonumber
\end{eqnarray}
where ${\bf K}(x)$ denotes the elliptic integral of the first kind.
Here we made use of a formula\cite{InteTable};
\begin{eqnarray}
\displaystyle
\int_{u}^{b}
{dx\over \sqrt{(a-x)(b-x)(x-c)(x-d)}}\nonumber\\
={2\over \sqrt{(a-c)(b-d)}}{\bf F}[\sin^{-1}\sqrt{(a-c)(b-u)\over(b-c)(a-u)},
\sqrt{(b-c)(a-d)\over(a-c)(b-d)}]
\nonumber
\end{eqnarray}
 and the relation ${\bf F}({\pi/2},q)={\bf K}(x)$
where
${\bf F}(\kappa,x)$ denotes the elliptic integral of the second kind.
Here $a>b>u\geq c>d$.

\section*{Appendix B}
\subsection*{Bethe-Salpeter Equation for particle-hole Channel}
Here we present the derivation of the solution of the Bethe-Salpeter
equation for particle-hole channel (\ref{eqn:integral eq for the ladder
vertex}).
The equation to be solved is
\begin{equation}
\Gamma(k+q,k)=1+{1\over2}{T\over N}\sum_{k'}J_{\k-\k'}{\cal
G}(k'+q)\Gamma(k'+q,k'){\cal G}(k').
\label{eqn:BS Eq}\end{equation}
By noting that
\begin{equation}
J_{\k-\k'}=2J(\cos k_{x}a\cos k'_{x}a+\sin k'_{x}a\sin k'_{x}a+\cos k_{y}a\cos
k'_{y}a+\sin k'_{y}a\sin k'_{y}a) ,
\end{equation}
we can rewrite (\ref{eqn:BS Eq}) in the form,
\begin{equation}
\Gamma(k+q,k)=1+J C(q) \,\,\,(\cos k_{x}a+\cos k_{y}a)+J S(q)\,\,\, (\sin
k_{x}a+\sin k_{y}a).
\end{equation}
Here
\begin{equation}
C(q)={1\over 2}(\overline{\cos k'_{x}a+\cos k'_{y}a})\,\,,\,\,\,S(q)={1\over
2}(\overline{\cos k'_{x}a+\cos k'_{y}a}),
\label{eqn:SC}\end{equation}
where
\begin{equation}
{\overline{(\cdots\cdots)}}\equiv
{T\over N}\sum_{k'}(\cdots\cdots){\cal G}(k'+q)\Gamma(k'+q,k'){\cal G}(k'),
\end{equation}
and $C(q)$ and $S(q)$  are to be determined self-consistently.
By inserting (\ref{eqn:SC}) back into (\ref{eqn:BS Eq}), we obtain the
following equations,
\begin{eqnarray}
C(q)&=&-{1\over 2}\chi_{1}(q)-{1\over 2}JC(q)\chi_{3}(q)-{1\over
2}JS(q)\chi_{4}(q),\\
S(q)&=&-{1\over 2}\chi_{2}(q)-{1\over 2}JC(q)\chi_{4}(q)-{1\over
2}JS(q)\chi_{5}(q).
\end{eqnarray}
where $\chi_{1},\cdots,\chi_{5}$ have been defined in (\ref{eqn:weighted
bubble}).
Finally by solving these simultaneous equations, we obtain (\ref{eqn:simeq C})
and (\ref{eqn:simeq S}).


\section*{Figure Captions}
\baselineskip 20 pt
\smallskip
Fig. 1: From left to right, the energy contour, corresponding DOS, and the
structure
of the saddle point near the Y-point for  \\
(a) $\alpha=0$,\\
(b) $\alpha=0.5$, \\
(c) $\alpha=1$.
The thick line in each figure represents the Fermi contour at the half filling
and
the broken line corresponds to the energy level at which the DOS becomes
divergent.

\smallskip
\noindent
Fig. 2: The fundamental processes induced by $J_{\q}$.\\
(a) The scattering process  with spin flip (type-a),\\
(b) the scattering processes between parallel spins (type-b) and \\
(c) the scattering processes between anti-parallel spins (type-c).
Here the straight line and the wavy line represent respectively
the green's function of an  itinerant carrier and the antiferromagnetic
interaction.

\smallskip
\noindent
Fig. 3: The string type summation for the transverse susceptibility.

\smallskip
\noindent
Fig. 4: Bethe-Salpeter equation for the triangle vertex.

\smallskip
\noindent
Fig. 5: The process corresponding to $\tilde\chi_{0 (n)}^{\sigma\sigma'}$.

\smallskip
\noindent
Fig. 6: The process corresponding to (\ref{eqn:n-th C}) in the text.

\smallskip
\noindent
Fig. 7: Temperature dependence of $X(T)$ in the self energy for\\
(a) $\alpha=0$; $n=0.85$ and\\
(b) $\alpha=1$; $n=0.85$.
We set $J=t/4$.

\smallskip
\noindent
Fig. 8: Scan of $\chi_{0}$, $\cdots$, $\chi_{5}$
along the $\Gamma\rightarrow X\rightarrow M \rightarrow\Gamma$ line
 in the first Brillouin zone for\\
(a) $\alpha=0$; $n=0.75$ and\\
(b) $\alpha=1$; $n=0.75$.
We set $T=0.04t\sim 60$[K]

\smallskip
\noindent
Fig. 9: $J C(q) \chi_{1}(q)$ and $J S(q) \chi_{2}(q)$ in the
first Brillouin zone for\\
(a) $\alpha=0$; $n=0.95, 0.75$ and\\
(b) $\alpha=1$; $n=0.95, 0.75$.

\smallskip
\noindent
Fig. 10: $\chi_{0}(q)$ and $\tilde\chi(q)$ in the
first Brillouin zone for\\
(a) $\alpha=0$; $n=0.95, 0.75$ and\\
(b) $\alpha=1$; $n=0.95, 0.75$.
\end{document}